\documentclass[aps,prl,twocolumn,groupedaddress,nofootinbib]{revtex4-2}

\usepackage{physics}
\usepackage{amsmath,amssymb,bm}
\usepackage{dcolumn}
\usepackage{xcolor}
\usepackage{graphicx}
\usepackage{times}

\begin{document}

\title{Irreversibility and Entropy Exclusion in Collisionless Plasmas}

\author{A. G. Tevzadze}
\affiliation{Evgeni Kharadze Georgian National Astrophysical Observatory, Abastumani 0301, Georgia}
\affiliation{Faculty of Exact and Natural Sciences, Javakhishvili Tbilisi State University, Tbilisi 0179, Georgia}

\begin{abstract}
We examine entropy production in reduced descriptions of collisionless plasmas. Introducing a closure dependent entropy hierarchy, we show that non-conservative moment closures generate a residual entropy associated with irreversible information loss, whereas invariant closures remain reversible. The monotonic growth of this residual entropy imposes a statistical realizability constraint on macroscopic plasma states, excluding regions of phase space independent of dynamical stability. For anisotropic plasmas, we evaluate entropy production within a second-order moment closure, identifying contributions from transport and magnetic field inhomogeneity. The resulting entropy exclusion boundary is broadly consistent with observed anisotropy distributions in space plasmas. Statistical realizability thus emerges as an organizing principle for reduced collisionless plasma descriptions.
\end{abstract}

\maketitle

Collisionless plasmas are commonly described as dynamically reversible systems whose macroscopic behavior is constrained primarily by conservation laws and dynamical stability. In this work, we show that this picture is incomplete. Even in the absence of collisions or linear instabilities, reduced descriptions of collisionless plasmas generically produce irreversible entropy through the loss of unresolved phase-space structure. Consequently, dynamical stability alone is insufficient to characterize accessible plasma states: statistical realizability provides an independent and complementary constraint on macroscopic phase space.

This apparent tension reflects the distinction between formal microscopic reversibility and physical irreversibility, long emphasized in statistical descriptions of collisionless plasma turbulence \cite{Krommes1993}. Although Vlasov dynamics is formally time reversible, the physical evolution of collisionless plasmas is generically irreversible owing to phase space filamentation, coarse graining, microturbulence, and other nonequilibrium effects, as observed in kinetic turbulence studies of weakly collisional plasmas \cite{Howes2008}. In reduced descriptions, this irreversibility cannot be eliminated but is instead displaced to higher, unresolved orders of the infinite moment hierarchy, consistent with earlier analyses of collisionless phase mixing and reduced kinetic models \cite{Dorland1993}. Conservative closures preserve reversibility at a given order by deferring entropy production to higher moments; from this perspective, non-conservative closures provide a self-consistent framework for making entropy production explicit within reduced collisionless plasma models.

Motivated by this viewpoint, we introduce a closure dependent entropy hierarchy that distinguishes between reversible and irreversible moment closures and demonstrate how entropy production arises in reduced collisionless plasma models. Focusing on anisotropic plasmas, we derive the entropy production mechanism at second order and obtain an explicit realizability constraint that excludes regions of macroscopic phase space. We illustrate the resulting exclusion principle through comparison with representative space plasma observations and conclude by discussing its implications for reduced descriptions of collisionless plasma flows.

To establish the formal framework underlying these results, we begin with the kinetic description of a collisionless plasma. The kinetic state of the plasma may be described by the single particle distribution function
$f(\bm{r},\bm{v},t)$, with species indices omitted for brevity. Dynamics of collisionless plasma may
be described by the Vlasov equation:
\begin{equation}
\frac{\partial f}{\partial t}
+ \bm{v}\cdot \frac{\partial f}{\partial \bm{r}}
+ \frac{\bm{F}_{\mathrm L}}{m}\cdot \frac{\partial f}{\partial \bm{v}} = 0 ~,
\label{eq:vlasov}
\end{equation}
where $\bm{F}_{\mathrm L} = q \left( \bm{E}+\bm{v}\times \bm{B}/c \right)$ is the Lorenz force.
Fluid description can be obtained by taking velocity moments of the distribution function of
the different order. The bulk density and velocity of the fluid model can be defined as:
\begin{equation}
\rho = m\int f\dd^3v ~,
\qquad
u_i=\frac{m}{\rho}\int v_i f\dd^3v ~.
\label{eq:rho_u}
\end{equation}
Higher order moments can be described using $n$-th order central velocity integrals:
\begin{equation}
M^{(n)}_{i_1\cdots i_n} = m\int c_{i_1}\cdots c_{i_n} f \dd^3v ~,
\label{eq:Mn_def}
\end{equation}
where $c_i=(v_i-u_i)$. Taking successive velocity moments of the Vlasov equation yields an
infinite hierarchy of evolution equations, in which the equation for the $n$-th moment couples
to the $(n+1)$-th moment (see BBGKY hierarchy \cite{Bogo1946} and its moments formulation \cite{Grad1949}). In compact operator form, this hierarchy can be written schematically as:
\begin{equation}
\frac{\dd}{\dd t} M^{(n)} + \mathcal{L}^{(n)}M^{(n)} + \mathcal{B}^{(n)}M^{(n)}
= -\nabla\cdot M^{(n+1)} ~,
\label{eq:hierarchy_compact}
\end{equation}
where $\mathcal{L}^{(n)}$ represents phase-space deformation due to bulk velocity gradients
(shear, strain, and compression), while $\mathcal{B}^{(n)}$ accounts for the action of the magnetic
field through fast gyromotion, where $\dd/\dd t=\partial/\partial t+(\bm{u}\cdot\nabla)$.
To obtain a finite system of equations, the infinite moment hierarchy must be closed. In turn, closure
models may be classified by the number of velocity moments they retain.

Moment closures may be grouped into two structural classes. In what follows, we refer to closures that enforce exact invariants or vanishing higher moments $(M^{(n+1)}=0)$ as \emph{hard} closures, while closures that prescribe higher order moments without evolving them dynamically are termed \emph{soft} closures. This distinction is consistent with entropy and realizability based classifications of moment closures \cite{Levermore1996}, as well as with Hamiltonian formulations of reduced models \cite{Morrison1998}.

Although the Vlasov equation is strictly time reversible, real collisionless plasmas exhibit irreversible macroscopic behavior. Each moment closure implicitly selects a distinct statistical ensemble of macroscopic plasma states by prescribing how unresolved kinetic information is retained or discarded. The distinction between hard and soft closures is therefore structural rather than quantitative: even arbitrarily accurate soft closures generically permit irreversible information leakage unless all higher moments are evolved self-consistently. This does not contradict the underlying reversible dynamics. Vlasov evolution transfers information to progressively finer phase space scales through filamentation. In real systems, however, observations and models are limited to finite resolution and reduced observables. Any such reduced description, whether through coarse graining, finite resolution, or moment closure irreversibly discards this fine-scale information, converting formally reversible microscopic dynamics into effective macroscopic entropy production (see e.g., \cite{EyinkSreenivasan2006,Schekochihin2008}).

In the present framework, irreversibility is not a consequence of dynamical instability, chaos, or dissipation, but arises structurally from the reduced representation of the collisionless plasma. Soft moment closures allow unresolved kinetic degrees of freedom to influence the dynamics without being evolved self-consistently, leading to irreversible information loss even though the underlying Vlasov dynamics remain time reversible. Irreversibility is therefore imposed by the structure of the closure itself, rather than by the microscopic equations of motion.

As an illustrative example of the link between closure and reversibility, we consider a hard 1st order closure corresponding to a cold, pressureless fluid, which is reversible. In contrast, a soft 1st order closure introduces a 2nd order velocity moment, pressure, normally associated with collisional effects, without evolving it dynamically, leading to irreversible entropy production despite the collisionless dynamics. The same mechanism generalizes to higher-order soft closures, in which entropy growth originates from explicit forcing by unresolved $(n+1)$-th moments.

We define $f^{(n)}$ as the $n$-moment representative distribution, which reproduces all velocity
moments $M^{(k)}$ with $k\le n$ and provides a closure for the fluid model. The specific form of
$f^{(n)}$ reflects the particular realization of the closure at order $n$, with a natural choice given by
the maximum-entropy realization. We then define the $n$-th order entropy associated with a given
closure level by
\begin{equation}
S^{(n)} = -\int f^{(n)} \ln f^{(n)} \dd^3v ~.
\label{eq:Sn_def}
\end{equation}
Note, that the sequence of $n$-order entropies obeys the entropy ordering inequality
\begin{equation}
S^{(n+1)} \le S^{(n) ~},
\label{eq:ordering}
\end{equation}
reflecting the reduction of uncertainty as higher order velocity moments are resolved. In these
notations the exact Gibbs--Boltzmann entropy may be recovered in the asymptotic limit:
\begin{equation}
S = \lim_{n\to\infty} S^{(n)} ~.
\label{eq:limit_entropy}
\end{equation}

We refer to a system as soft ``leaky moment'' closure, if higher order velocity moments
influence the dynamics without being evolved self-consistently, allowing irreversible information transfer
out of the resolved moment subspace. At a fixed moment order $n$, the entropy $S^{(n)}$ is not uniquely
defined but depends on the choice of closure. We therefore distinguish between a reversible entropy
$S^{(n)}_0$, associated with hard (invariant) closures, and an irreversible entropy $S^{(n)}_1$,
corresponding to soft (non-conservative) closures that prescribe higher order moments without evolving them
dynamically. While $S^{(n)}_0$ is conserved, $S^{(n)}_1$ naturally exhibits entropy production due to
irreversible information leakage into unresolved kinetic degrees of freedom. The subscripts $0$ and $1$
denote, respectively, hard and soft realizations of the $n$-order entropy:
\begin{equation}
\frac{\dd}{\dd t}S^{(n)}_0 = 0 ~,
\qquad
\frac{\dd}{\dd t}S^{(n)}_1 \ge 0 ~.
\label{eq:S0S1}
\end{equation}
Hence, we introduce a residual entropy contribution that quantifies the information associated with
irreversibility at the $n$-th order closure:
\begin{equation}
R^{(n)} = S^{(n)}_1 - S^{(n)}_0 ~.
\label{eq:Rn_def}
\end{equation}
The monotonic growth of the residual entropy
\begin{equation}
\frac{\dd}{\dd t}R^{(n)} \ge 0
\label{eq:Rn_monotone}
\end{equation}
has an important consequence: it imposes an additional realizability constraint on the phase space of
macroscopic variables generated by a given closure. This constraint is independent of, and complementary
to, the standard realizability requirements such as positivity of the distribution function, positivity of the pressure tensor, and dynamical realizability of the moment system. Since $R^{(n)}$ quantifies unresolved kinetic structure, only those macroscopic states that can be reached through monotonic information loss are
dynamically admissible. States that would require a decrease of $R^{(n)}$ are therefore forbidden within
the reduced description. In this sense, entropy production in soft closures does not merely signal
irreversibility, but may actively restrict the realizable phase space region associated with a particular
closure.

As an illustrative example, irreversibility in the first order MHD closure may be associated with the 
entropy functional
\begin{equation}
\frac{\dd}{\dd t} R^{(1)} = \frac{\dd}{\dd t}\ln \left(\frac{P}{\rho^\gamma}\right) \ge 0,
\label{eq:R1}
\end{equation}
where $\gamma$ is the adiabatic index used in the MHD closure. This limit is degenerate, since a strictly conservative first order closure corresponds to a cold fluid with vanishing pressure. The associated hard closure entropy $S^{(1)}_0$ is therefore trivial and conserved, reflecting the absence of internal kinetic degrees of freedom. Introducing pressure at first order constitutes a soft closure, so the residual entropy growth condition reduces to the standard entropy production constraint and does not impose an additional realizability restriction. This degeneracy is specific to first order closures and is not expected to persist for higher order moment systems, where genuinely new entropy based constraints emerge.

Since the first order case is degenerate and yields no additional realizability constraint, we turn to 2nd order closures, where the monotonic growth of residual entropy may lead to a nontrivial entropy exclusion condition. We evaluate the 2nd order residual entropy, $R^{(2)}$, using the Chew–Goldberger–Low (CGL) model \cite{CGL1956} as a second-order hard closure, with heat fluxes incorporated as a soft closure correction. The parallel and perpendicular pressure components then obey the following soft closure model equations:
\begin{align}
\frac{\dd}{\dd t}\left(\frac{P_{\parallel}B^2}{\rho^3}\right)
&=
-\frac{B^3}{\rho^3}
\left[
\nabla_{\parallel}\!\left(\frac{Q_{\parallel}}{B}\right)
+\frac{2Q_{\perp}}{B^2}\nabla_{\parallel}B
\right] ~,
\label{eq:CGL_soft_par}
\\
\frac{\dd}{\dd t}\left(\frac{P_{\perp}}{\rho B}\right)
&=
-\frac{B}{\rho}\nabla_{\parallel}\!\left(\frac{Q_{\perp}}{B^2}\right) ~,
\label{eq:CGL_soft_perp}
\end{align}
where $Q_{\parallel}$ and $Q_{\perp}$ are the parallel and perpendicular heat flux components.
For brevity, we use the field aligned derivative:
\begin{equation}
\nabla_{\parallel}\equiv \frac{1}{B}\left(\bm{B}\cdot\nabla\right) ~.
\label{eq:par_deriv}
\end{equation}
Here, the anisotropic pressure components $P_{\parallel}$ and $P_{\perp}$ are derived from the 2nd order velocity moments, whose
evolution is coupled to 3rd order moments through the heat fluxes $Q_{\parallel}$ and $Q_{\perp}$. In the
classical hard CGL closure ($Q_{\parallel}=Q_{\perp}=0$), Eqs.~\eqref{eq:CGL_soft_par} and
\eqref{eq:CGL_soft_perp} reduce to the conservative form of the two CGL adiabatic invariants.
Invoking CGL entropy form
\begin{equation}
S_{\mathrm{CGL}}=\frac{1}{3}\ln\!\left(\frac{P_{\parallel}P_{\perp}^2}{\rho^5}\right) ~,
\label{eq:SCGL}
\end{equation}
in the hard CGL closure derivative of this form is zero, ($S^{(2)}_0=S_{\mathrm{CGL}}$) while in the soft
closure using the same fluid variables we may define time derivative of the irreversible entropy as follows:
\begin{equation}
\frac{\dd}{\dd t}S^{(2)}_1
=\frac{\dd}{\dd t}\ln\!\left(\frac{P_{\parallel}P_{\perp}^2}{\rho^5}\right)^{1/3} ~.
\label{eq:S2_1_def}
\end{equation}
Now, using Eqs.~\eqref{eq:S0S1}, \eqref{eq:Rn_def}, \eqref{eq:CGL_soft_par}, \eqref{eq:CGL_soft_perp}
we may derive dynamical equation for the 2-nd order entropy residual form:
\begin{equation}
\frac{\dd}{\dd t}R^{(2)} = \Sigma_{\parallel}+2\Sigma_{\perp}+2\Sigma_{B} ~,
\label{eq:R2_sources}
\end{equation}
with
\begin{align}
\Sigma_{\parallel} &=
-U_{\parallel} \nabla_{\parallel}\ln \left(\frac{Q_{\parallel}}{B}\right)^{1/3} ~,
\label{eq:Sig_par}\\
\Sigma_{\perp} &=
-U_{\perp} \nabla_{\parallel}\ln \left(\frac{Q_{\perp}}{B^2}\right)^{1/3} ~,
\label{eq:Sig_perp}\\
\Sigma_{B} &=
-U_{\perp} \frac{P_{\perp}}{P_{\parallel}}
\nabla_{\parallel}\ln(B)^{1/3} ~,
\label{eq:Sig_B}
\end{align}
where we have introduced the characteristic anisotropic heat flux transport velocities:
\begin{equation}
U_{\parallel} = {Q_{\parallel} / P_{\parallel}} ~,
\qquad
U_{\perp} = {Q_{\perp} / P_{\perp}} ~.
\label{eq:Udefs}
\end{equation}
Importantly, the form of Eq. \eqref{eq:R2_sources} is independent of the specific realization of the soft closure; only the explicit evaluation of the exclusion boundary requires a closure prescription.
It reveals physical picture of the entropy transfer sources that
contribute to the overall entropy production in the system that include contributions from parallel heat flux ($\Sigma_{\parallel}$), perpendicular heat flux ($\Sigma_{\perp}$) and field aligned gradient of the magnetic field strength ($\Sigma_{B}$). Interestingly, the magnetic field inhomogeneity acts as an effective sink of macroscopic information by directly entering the entropy budget, rather than influencing entropy production solely through pressure anisotropy or transport.
Hence, Eq. (\ref{eq:R2_sources}) sets the realizability condition in the phase space:
\begin{equation}
\Sigma_{\parallel}+2\Sigma_{\perp}+2\Sigma_{B} \geq 0 ~. \label{Real}
\end{equation}
To derive a concrete phase-space constraint from Eq. (\ref{Real}), one must specify the relevant phase space variables and adopt a soft closure prescription, i.e., prescribe how the parallel and perpendicular heat fluxes $Q_{\|}$ and $Q_{\perp}$ are determined. For this purpose, we introduce the pressure anisotropy 
$\alpha$ and the plasma-$\beta$ parameter, widely employed to quantify the thermodynamic state of collisionless plasma flows inferred from observations:
\begin{equation}
\alpha = {P_{\perp} / P_{\|}} ~,~~ \beta_{\parallel} = 8 \pi {P_{\|} / B^2} ~.
\end{equation}
For further calculations we adopt an anisotropic flux limited closure, in which the parallel heat flux is limited by the parallel thermal (sound) speed, consistent with collisionless free streaming arguments \cite{Hammett1990}, while the perpendicular heat flux is limited by the Alfv\'en speed, reflecting magnetic constraints on cross field transport \cite{Parker1965}:
\begin{equation}
U_{\|} = \chi \left( {P_{\|} \over \rho} \right)^{1/2} ~,~~
U_{\perp} = \chi \left( {B^2 \over 4 \pi \rho} \right)^{1/2} ~,
\end{equation}
where $\chi<1$  is a dimensionless flux-limiter parameter accounting for the regulation of heat flux transport by collisionless plasma dynamics. For qualitative estimates we adopt a weakly varying field aligned profile $\nabla_{\|} P_{\|} \approx 0$ and $\nabla_{\|} \rho \approx 0$. This reduces Eq. (\ref{Real}{}) to an envelope condition that defines the exclusion boundary function $\alpha = \alpha(\beta)$:
\begin{equation}
{{\rm d} \ln \alpha \over {\rm d} \ln \beta_{\parallel} } + {1 \over 2} \left[ 1 - \alpha + \left({\beta_{\parallel} \over 8}\right)^{1/2} \right] \leq 0 ~. \label{Disp}
\end{equation}
which constitutes a differential inequality in the $(\alpha,\beta)$ plane. This inequality restricts the admissible directions of phase space trajectories generated by the reduced dynamics. The equality case defines an envelope that cannot be crossed by realizable trajectories: macroscopic states lying below this boundary would require negative entropy production and are therefore dynamically inaccessible within the reduced description. We refer to this constraint as a phase space restriction principle, or entropy exclusion, which forbids access to certain macroscopic states.

Equation (\ref{Disp}) is solved numerically to obtain the corresponding entropy exclusion envelope in the $(\beta_{\parallel},\alpha)$ plane for parameters representative of the solar wind. As shown in Fig. 1, the resulting entropy excluded region closely coincides with the observed anisotropy distribution, particularly in the low-$\beta_{\parallel}$ regime. The agreement is notable given the minimal nature of the soft closure and suggests that entropy exclusion provides a natural explanation for the systematic absence of fluctuations at low $\beta_{\parallel}$.  The entropy-exclusion boundary should be interpreted as an envelope approached asymptotically by macroscopic trajectories under irreversible evolution, rather than as a sharp threshold dynamically crossed or reflected from.

\begin{figure*}[t]
\centering
\includegraphics[width = 0.9\textwidth]{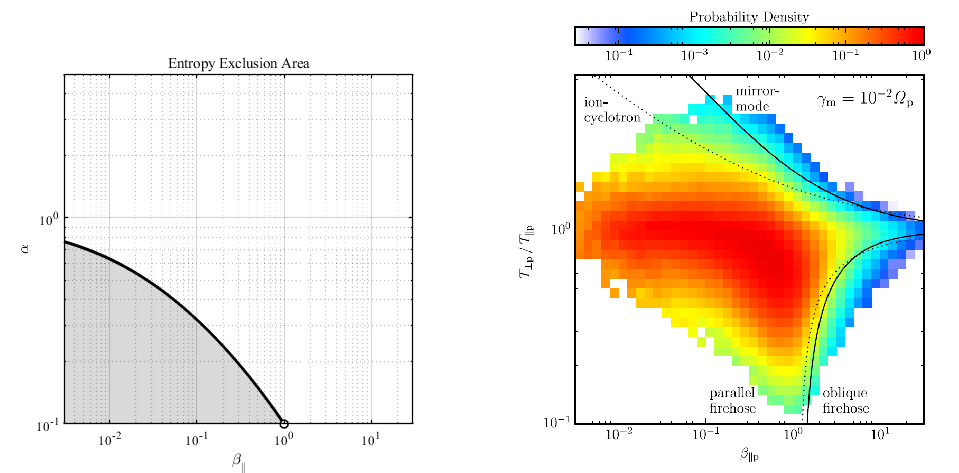}
\caption{
Entropy-admissible set in anisotropic plasma states.
Left panel: Boundary of the entropy exclusion area in $(\beta_{\parallel},\alpha)$.
The shaded region is excluded by the constraint set by the residual entropy growth argument.
Right panel: Solar wind distribution in $(\beta_{\parallel p},T_{\perp p}/T_{\parallel p})$
adapted from Ref.~\cite{Klein2019Entropy} (CC BY 4.0).
}
\label{fig1}
\end{figure*}

We have shown that irreversibility and entropy production in collisionless plasmas imply the existence of excluded macroscopic states in phase space. In the present framework, irreversibility is structural: it arises from reduced representations of the plasma rather than from dynamical instability or dissipation in the underlying Vlasov dynamics. Using a 2nd order soft moment closure, we demonstrated that irreversible information loss to unresolved kinetic scales generates a residual entropy whose monotonic growth enforces an exclusion principle. As a result, even dynamically stable collisionless plasma states may be statistically forbidden.

This constraint is not purely thermodynamic. Magnetic field inhomogeneity enters explicitly into the entropy budget, placing magnetic structure on equal footing with pressure anisotropy and transport. Entropy production therefore acts not merely as a diagnostic of irreversibility, but as an active mechanism restricting access to macroscopic phase space. The excluded states identified here are not associated with linear micro-instabilities or marginal stability thresholds. Instead, statistical exclusion imposes a global, cumulative constraint arising from irreversible information loss. Whereas instability thresholds restrict plasma states through local growth rates, statistical realizability limits the very existence of admissible macroscopic states within reduced descriptions, providing a weaker but more universal criterion governing accessible configurations of collisionless plasma flows.

A comprehensive analysis of residual entropy production across all closures and higher order moment systems is beyond the scope of this work. Our aim is instead to establish, at a conceptual level, that monotonic residual entropy growth generically imposes entropy exclusion constraints on macroscopic phase space in collisionless plasmas. By demonstrating this mechanism for a concrete 2nd order closure and showing consistency with representative observations, we illustrate the approach without attempting an exhaustive survey of closure models. The primary novelty of this work lies in identifying entropy exclusion as an organizing principle for reduced collisionless plasma descriptions.

More broadly, these results suggest that entropy driven statistical realizability imposes a generic constraint on reduced descriptions of collisionless systems, with implications for model validation, turbulence, and macroscopic stability. In turbulent plasmas, statistical exclusion may act as a global constraint shaping long time macroscopic states, independent of local cascade dynamics or instability driven saturation. More generally, such constraints may be intrinsic to reduced descriptions of collisionless Hamiltonian systems whenever fine-scale phase space information is discarded.

\bibliographystyle{apsrev4-2}

\end{document}